\documentstyle[12pt]{article}
\newcommand{\square}{\kern1pt\vbox{\hrule height 1.2pt\hbox{\vrule width
1.2pt\hskip 3pt
   \vbox{\vskip 6pt}\hskip 3pt\vrule width 0.6pt}\hrule height 0.6pt}\kern1pt}
\newcommand{\beq}{\begin{equation}}
\newcommand{\eeq}{\end{equation}}
\newcommand{\barr}{\begin{eqnarray}}
\newcommand{\earr}{\end{eqnarray}}
\newcommand{\nn}{\nonumber\\}
\newcommand{\lmn}{_{\mu\nu}}

\newcommand{\uab}{^{\alpha\beta}}
\newcommand{\epon}{\epsilon_1}
\newcommand{\eptw}{\epsilon_2}
\newcommand{\epth}{\epsilon_3}
\newcommand{\eptt}{\epsilon_{23}}

\begin{document}
\renewcommand{\Large}{\normalsize}
\renewcommand{\huge}{\normalsize}
\newcommand{\lsim}{\mbox{\raisebox{-1.ex}
{$\stackrel{\textstyle<}{\textstyle \sim}$}}}
\newcommand{\grle}{\mbox{\raisebox{-.7ex}
{$\stackrel{\textstyle>}{\textstyle<}$}}}
\newcommand{\legr}{\mbox{\raisebox{-.7ex}
{$\stackrel{\textstyle<}{\textstyle>}$}}}
\newcommand{\gsim}{\mbox{\raisebox{-1.ex}
{$\stackrel{\textstyle >}{\textstyle \sim}$}}}

\begin{titlepage}
\baselineskip .15in
\begin{center}
{\bf

\vskip 1.5cm
{\large Multiple Field Scalar-Tensor Theories \\[1em]
 of Gravity and Cosmology}

}\vskip .8in

{\sc  Andrew L. Berkin }$^{(*)}$\\[1em]
and {\sc Ronald W. Hellings }$^{(**)}$\\[2em]
 {\em Jet Propulsion Laboratory \\
California Institute of Technology \\
4800 Oak Grove Drive \\
Pasadena, CA 91109}\\

\end{center}
\vfill
\begin{abstract}
   We consider multiple scalar fields coupled to gravity, with special
attention given to two-field theories. First, the conditions necessary for
these theories to meet solar system tests are given. Next, we investigate
the cosmological evolution of the fields to see if these conditions can be met.
Solutions are found in the dust era, as well as radiation and cosmological
constant dominated epochs. The possibility of inflation in these theories
is discussed.
While power law growth of the scalar fields can yield the appropriate
conditions to meet solar system constraints, these solutions are unstable.
\end{abstract}

\vfill
\begin{center}
October, 1993
\end{center}
\vfill
(*)~~Email: berkin@krazy.jpl.nasa.gov\\
(**)~~Email: rwh@grouch.jpl.nasa.gov\\
\end{titlepage}

\baselineskip .3in

\begin{flushleft}
\section{Introduction}
\end{flushleft}
\vspace{.5 cm}
\normalsize
\baselineskip = 24pt

Scalar-tensor (ST) theories are alternative models of gravity which provide
a theoretical framework within which general relativity (GR) may be tested.
Many important tests so far have used the post-Newtonian approximation, and
have relied upon corrections to dynamics in the solar system. Such tests
force ST theories to limits where they greatly resemble GR. For example, the
current limit on Brans-Dicke theories\cite{JBD} requires
$\omega>500$\cite{BDlimit}, where
$\omega\rightarrow\infty$ recovers GR and $\omega\approx 1$ would be a
natural value to expect {\it a priori}.

   However, despite the strong resemblance to GR in our solar system,
gravity may be radically different in other regimes. The gravitational
field in our solar system is weak, severely limiting the parameter space
of gravity tested. Theories which differ
from GR in strong gravitational fields, but agree with solar system
constraints, are needed to further test GR.
Recently, a class of ST theories with multiple scalar fields has been
proposed\cite{Nfield}. These theories can satisfy the solar system criteria
to arbitrary accuracy, but still diverge from GR in other limits, for
example in the strong field regime around binary pulsars. Thus, such theories
provide an important test of GR in a previously sparsely tested regime.

   Aside from their importance as generalizations of standard ST theories,
multiple scalar field theories have a second motivation. Such additional
scalars coupled to gravity appear in Kaluza-Klein\cite{kk} and string
theories\cite{string} which seek to unify gravity with other forces. Tests
of the correctness of GR thus provide constraints on such theories.
Furthermore, ST theories combined with Grand Unified Theories can provide
adjustments to inflationary models of the early universe\cite{infl} which
allow the phase transition to complete in old inflation\cite{extend} and
remove the fine-tuning of new and chaotic models\cite{gets}.
The inflationary universe
not only solves several longstanding cosmological problems, but also provides
the only currently known source of seed density fluctuations which obey the
magnitude and spectrum of the microwave background found by COBE\cite{cobe}.
While these ``extended" versions of old
inflation are severely constrained by the solar system
tests\cite{wbg}, multiple ST models may offer successful conditions for
inflation which still obey all observational tests.
Thus, we seek the viability of inflation in multiple field models where
the solar system constraints may be avoided.

   In this paper we consider a particular set of multiple scalar ST theories
which are a straightforward generalization of nonminimally coupled models.
The one-field nonminimally coupled model is
related to the JBD theory by a redefinition of the scalar field.
Our starting action is thus
different from those used previously\cite{Nfield}, where the action was
taken after a conformal transformation had been made so that the gravity
sector appeared normal.
We initially consider a
model with an arbitrary number $N$ of scalar fields, and derive the
conditions necessary to meet the solar system constraints. These conditions
depend on the asymptotic values of the fields far from the solar system.
Previous work used the ansatz that these values were constants  which
satisfy all the equations. However, the actual values of these fields should
be set by their cosmological evolution.

In order to study their cosmological evolution,
we first find a generic set of
solutions which exhibit power law behavior, and give exact solutions
in the dust, radiation, and cosmological constant dominated eras. Next, we
specialize to a two-field model, which elucidates the main features of the
behavior. The stability of these solutions in time is then examined, and
small deviations are found to diverge from the exact power law behavior
as the universe expands. Hence, while solutions
of multiple field ST theories are possible which meet the solar system
constraints and field equations for cosmological evolution, they are unstable.
This finding calls into question the ability of these models to provide
a true test of GR. We close with comments on the possibilities of success
of these theories and their potential role in inflation.

\setcounter{equation}{0}
\begin{flushleft}
\section{$N$ Field Models}
\end{flushleft}
\vspace{.5 cm}
\normalsize
\baselineskip = 24pt

   We begin by considering the action
\beq
S=\int d^4x\sqrt{-g}\left(f(\phi)R-G_{AB}g^{\alpha\beta}\phi_{A,\alpha}
\phi_{B,\beta}-2U(\phi)+16\pi L_m\right),\label{eq1}
\eeq
where $\phi$ is an $N$ component scalar field whose individual components
will be denoted with capital Roman letters, $U$ is a potential, $G_{AB}$
describes the kinetic coupling of the fields, $L_m$ is the
Lagrangian for other matter and summation over $A$ and $B$ is implied.
The scalar field enters the Lagrangian
with metric coupling to matter.
By an appropriate linear transformation
and rescaling of the $\phi$, $G_{AB}$ may be taken to be diagonal with
entries of $\pm 1$ for each scalar field.
All other conventions are identical to
those of Misner, Thorne and Wheeler\cite{mtw}. $f(\phi)=G\equiv 1/\kappa^2$,
where $G$ is Newton's constant,
recovers GR with scalar fields, while $f(\phi)=\xi\phi^2$ with $\phi$ a
singlet is the standard nonminimally coupled model. Making the field
redefinition $\Phi=\xi\phi^2$ for singlet $\phi$ places the action into
standard Brans-Dicke form.

   Varying this action with respect to the metric gives the gravitational
field equations
\beq
\frac12 g_{\mu\nu}\left(-fR+G_{AB}g^{\alpha\beta}\phi_{A,\alpha}\phi_{B,\beta}
+2U\right)-f_{;\mu\nu}+g_{\mu\nu}\square f+fR_{\mu\nu}
-G_{AB}\phi_{A,\mu}\phi_{B,\nu} = 8\pi T\lmn,\label{eq2}
\eeq
with $T\lmn$ the energy-momentum tensor corresponding to $L_m$. The trace
of this equation is
\beq
-fR+G_{AB}g^{\alpha\beta}\phi_{A,\alpha}\phi_{B,\beta}+4U+3\square f = 8\pi T.
\label{eq3}
\eeq
Varying with respect to $\phi$ gives the scalar field equations
\beq
f_{A}R+2G_{AB}\square\phi_B-2U_{A} = 0,\label{eq4}
\eeq
with a subscript $A$ referring to the partial derivative
$\partial/\partial\phi_A$.
In order to be viable, metric theories such as these must satisfy the
solar system experimental constraints. Of these, we concentrate in the
next section on the effects arising from first order space curvature
(PPN parameter $\gamma$).

\subsection{solar system constraints}

   To lowest order in the post-Newtonian approximation, the metric around
a test body of mass $M$ in the solar system is\cite{mtw}
\beq
ds^2=-\left(1-{2M\over r}\right)dt^2
+\left(1+2\gamma{M\over r}\right)[dx^2+dy^2+dz^2],
\label{eq5}
\eeq
where $\gamma=1$ recovers the value in GR. In the JBD theory,
$\gamma = (1+\omega)/(2+\omega)$ and current tests from the Viking lander give
$\omega>500$\cite{BDlimit}.

   We will use the $(00)$ and trace equations to determine the expression
for $\gamma$ in the theory of equation (\ref{eq1}). Note that
\beq
\square f = g\uab(f_{AB}\phi_{A,\alpha}\phi_{B,\beta}
+f_{A}\phi_{A;\alpha\beta})\approx f_{A}\square\phi_A.\label{eq6}
\eeq
To obtain this last approximation, we removed the first term by using the
fact that
$\nabla\phi={\cal O}(M)$ and keeping only lowest order in $M$.
Then, making use of equation (\ref{eq4}), we find
\beq
\square f=-\frac12G^{-1}_{AB}f_{A}f_{B}R+U_{A}f_{B}G^{-1}_{AB}.\label{eq7}
\eeq

The matter source for the solar system bodies will be taken to be pressureless
dust, with delta function stress-energy terms.
Making use of the representation of the delta function by $\nabla^2(1/r)$,
we therefore write
\beq
T_{00}=-{1\over4\pi}\nabla^2\left({M\over r}\right),\hspace*{1cm}
T_{ij}=0,\hspace*{1cm}
T={1\over4\pi}\nabla^2\left({M\over r}\right).\label{eq8}
\eeq
Again, to lowest order, the necessary curvature components are
\beq
R_{00}=-\frac12\nabla^2g_{00}=-\nabla^2\left({M\over r}\right),\hspace*{1cm}
R=2(1-2\gamma)\nabla^2\left({M\over r}\right).\label{eq9}
\eeq

   The $(00)$ equation then implies
\beq
-(f+f_{A}G^{-1}_{AB}f_{B})(1-2\gamma)\nabla^2\left({M\over r}\right)
+f\nabla^2\left({M\over r}\right)+U+U_{A}G^{-1}_{AB}f_{B}
=2\nabla^2\left({GM\over r}\right),
\label{eq10}
\eeq
where $G$ is Newton's constant. If we take the case $U=0$, which we will
consider throughout the rest of this paper, and define
\beq
C\equiv f_{A}G^{-1}_{AB}f_{B},\label{eq11}
\eeq
then we find
\beq
C-2\gamma(f+C)=-2G.\label{eq12}
\eeq
Similarly, the trace equation yields
\beq
(2f+3C)(1-2\gamma)=-2G.\label{eq13}
\eeq
Combining these two equations then gives
\beq
\gamma = {f+C\over f+2C}.\label{eq14}
\eeq
This theory will be identical to GR for solar system tests if $\gamma=1$,
which then implies
\beq
C\equiv f_{A}G^{-1}_{AB}f_{B}=0.\label{eq15}
\eeq
As long as the coupling satisfies this last relationship, multiple ST
theories can exactly replicate GR in the solar system
without being identical to GR. However, theories which satisfy (\ref{eq15})
may still be quite different in other gravitational regimes.
This behavior is in
contrast to single field ST gravity, where the solar system constraints force
the entire theory to be indistinguishable from GR.

   Of course, $\gamma$ need not be exactly unity, but need only
satisfy the current
experimental limits. Indeed, nature may dictate that $\gamma$ is not 1
but only some
value close to 1, in which case GR would fail. If this should be the case,
both single and
multiple ST theories could meet the new constraint with a $C$
that is a small but nonzero value, while higher order effects
would cause a larger discrepancy with GR in other regimes.

   Whether the condition of equation (\ref{eq15}) can be met depends on the
values of the scalar fields.
Because $G_{AB}$ may be diagonalized and rescaled, the vanishing of $C$
implies either the $f_A$ are all 0 in the solar system, or that some of
the scalars have negative kinetic coupling.
Also, if $C=0$, then equations (\ref{eq12})
and (\ref{eq13}) demand that $f$ be equal to the gravitational constant,
setting another condition.
For a large number of models, including the one here,
setting the background value of the fields to
zero certainly meets these conditions.
In many other models, such as that of \cite{Nfield},
values constant in time will suffice.
However, these
values are actually determined by the cosmological evolution of $\phi$,
to which we next turn our attention.

\subsection{cosmological evolution}

   For cosmology, we use a spatially flat Robertson-Walker universe, with
metric given by
\beq
ds^2=-dt^2+a^2(t)\left(dx^2+dy^2+dz^2\right),\label{eq16}
\eeq
where $a(t)$ is the time dependent scale factor. From the $(00)$ component
of the gravitational field equations (\ref{eq2}), we get
\beq
3H^2f-\frac12G_{AB}\dot{\phi}_A\dot{\phi}_B+3H\dot{f}-U=8\pi\rho,\label{eq17}
\eeq
where $H\equiv \dot{a}/a$ is the Hubble parameter, $\rho$ is the energy
density, and an overdot indicates differentiation with respect to time.
The trace equation gives
\beq
f(6\dot{H}+12H^2)+G_{AB}\dot{\phi}_A\dot{\phi}_B-4U+3\ddot{f}+9H\dot{f}
=8\pi(\rho-3P),\label{eq18}
\eeq
where $P$ is the pressure of matter, which is assumed to be a perfect fluid.
Finally, the scalar field equations, (\ref{eq4}), yield
\beq
\square\phi_A=-\frac12G^{-1}_{AB}f_{B}R+U_{B}G^{-1}_{AB}=0,\label{eq19}
\eeq
with the Ricci scalar given by
\beq
R=6\dot{H}+12H^2\label{eq20}
\eeq
One equation of the set (\ref{eq17}) - (\ref{eq19}) is extraneous due to the
symmetries of the spacetime and
the contracted Bianchi identities.

\noindent{\bf dust era}

   In the current dust dominated universe, with $U=0$, the constraint for
the solar system (\ref{eq15}) must be satisfied. Contracting equation
(\ref{eq19}) with $f_{A}$ gives
\beq
f_{A}\square\phi_A=-\frac12f_{A}G^{-1}_{AB}f_{B}R=0,\label{eq21}
\eeq
a rather strong
constraint on the behavior of the scalar fields in a dust universe.
The limits imposed by the null summation (\ref{eq21}) will be more
obvious when we consider a two field case in the next section.

   To better understand this constraint, we consider particular choices
for $f(\phi)$ of the form
\beq
f(\phi)=f_c+F_{AB}\phi_A\phi_B,\label{eq22}
\eeq
with $f_c$ and $F_{AB}$ constant. This model is a straightforward
generalization of the standard one field nonminimal coupling. There are no
added dimensional coupling constants. Only $f_c$, which
corresponds to a modified gravitational constant, has dimensions.

   Because these equations are nonlinear, general
solutions will be difficult, if not impossible, to obtain. We therefore
look for power law solutions, by making the ansatz
\beq
a(t)=a_c\left({t\over t_c}\right)^p,\hspace*{1cm}
\phi_A=b_A\left({t\over t_c}\right)^q,\label{eq23}
\eeq
where $a_c$, $t_c$, $b_A$, $p$, and $q$ are all constants.
Although this power law assumption restricts the generality of our
solutions, the problem does become tractable.
The dependence of the solar system constraints upon
cosmological evolution is manifested, a feature missing from just a
constant $\phi$ solution.
For GR, we know that the cosmological expansion is power law,
with $p$ being 2/3 and 1/2 in the dust and radiation regimes, respectively.
Thus, physically, our solutions allow easy comparison with the
standard cosmological behavior.

   With the above simplifications, equation (\ref{eq21}) gives
\beq
F_{AB}b_Ab_B{t^{2q-2}\over t_c^{2q}}(q^2-q+3pq)=0.\label{eq24}
\eeq
This equation is solved either by $q=0$, $q=1-3p$, or $F_{AB}b_Ab_B=0$.
The last condition causes $f$ to be constant, $f=f_c=G$. The gravitational
constant does not evolve.
The first condition, $q=0$, corresponds to
constant fields, while the middle condition relates the evolution of the
fields to that of the scale factor.

   We now examine the scalar field equations, (\ref{eq19}), which imply
\beq
b_A(q^2-q+3pq) = G^{-1}_{AB}F_{BC}b_C(-6p+12p^2).\label{eq25}
\eeq
If $q$ equals either 0 or $1-3p$, then the left hand side must be 0,
giving $p=0$ or $p=1/2$. The first case corresponds to a static universe,
in conflict with observation, and will be ignored. We thus retain only
$p=1/2$.

   Next consider the $(00)$ equation, which for power law behavior
becomes
\beq
{3p^2\over t^2}\left[f_c+F_d\left({t\over t_c}\right)^{2q}\right]
-{q^2G_d\over 2t^2}\left({t\over t_c}\right)^{2q}
+{6pqF_d\over t^2}\left({t\over t_c}\right)^{2q}
={8\pi\rho_d\over a_c^3}\left({t\over t_c}\right)^{-3p},\label{eq26}
\eeq
with the definitions
\beq
F_d\equiv F_{AB}b_Ab_B,\hspace*{1cm} G_d\equiv G_{AB}b_Ab_B\label{eq27}
\eeq
in the dust era. As noted from equation (\ref{eq25}), if $q=0$ or $q=1-3p$,
then $p=1/2$. However, plugging these values into equation (\ref{eq26}) shows
that the $(00)$ equation cannot be solved, because the powers of time will
not balance. Therefore, of the three possible solutions to equation
(\ref{eq24}), only the last one, $F_d=0$, has the possibility of being
consistent with both an expanding universe and the other field equations.

   Now, equation (\ref{eq25}) implies that
\beq
G_d={-6p+12p^2\over q^2-q+3pq}F_d = 0.\label{eq28}
\eeq
Using the fact that both $F_d$ and $G_d$ are zero, the $(00)$ equation
simplifies to
\beq
{3p^2\over\kappa^2t^2}={8\pi\rho_d\over a_c^3}\left({t\over t_c}\right)^{-3p},
\label{eq29}
\eeq
identical with the GR case. Thus, because of the cancellation necessary to
meet the solar system constraint (\ref{eq15}), the dust dominated universe
with power law expansion resembles GR in all cosmological aspects.
However, this does not mean that other cosmologies, such
as a radiation dominated universe, necessarily resemble GR, as we shall
see below.

\noindent{\bf radiation era}

   For a radiation dominated universe the energy density is
given by
\beq
\rho(t)=\rho_r a^4(t),\label{eq30}
\eeq
with $\rho_r$ a constant, and $P=\rho/3$. Since the current universe
is not radiation dominated, the solar system
constraint, equation (\ref{eq15}), need not apply.
Because the stress energy tensor for radiation is traceless, the trace
equation,
\beq
\left[{1\over\kappa^2}+F_r\left({t\over t_r}\right)^{2q}\right]
{(-6p+12p^2)\over t^2}+{q^2G_r\over t^2}\left({t\over t_r}\right)^{2q}
+{6q(2q-1)F_r\over t^2}\left({t\over t_r}\right)^{2q}
+{18pqF_r\over t^2}\left({t\over t_r}\right)^{2q}=0,\label{eq32}
\eeq
is most convenient to consider first.
Here $t_r$ is a constant, and $F_r$ and
$G_r$ correspond to the quantities (\ref{eq27}) in the radiation era.

   Balancing the powers of time then requires that $p=1/2$, exactly as
in the standard GR case. With this value, the first term in equation
(\ref{eq32}) vanishes, and the resulting time independent part is solved by
\beq
q=0,\hspace*{.5cm} {\rm or}\hspace*{.2cm} q={-3F_r\over G_r+12F_r}.\label{eq33}
\eeq
We then plug this result into the $(00)$ equation, which is identical with
equation (\ref{eq26}) except with a right hand side
\beq
{8\pi\rho_r\over a_r^4}\left({t\over t_c}\right)^{-4p}.\label{eq34}
\eeq

   If $q=0$, then the fields do not evolve, and
\beq
\frac34({1\over\kappa^2}+F_r)={8\pi\rho_r\over a_r^4}t_r^2.\label{eq35}
\eeq
This is similar to the Einstein GR case, except that the gravitational
constant is shifted by $F_r$.

   When $q=-3F_r/(G_r+12F_r)$, the $(00)$ equation that results is solved
by any one of three conditions. These are
\beq
\begin{array}{lcl}
F_r = 0 &\Rightarrow & q = 0\nn
G_r = 0 &\Rightarrow & q = -\frac14\label{eq36}\\
G_r = -6F_r &\Rightarrow & q = -\frac12.\nonumber
\end{array}
\eeq
This first case just reproduces the $q=0$ case above.
The latter two conditions cause cancellations such that
\beq
{3\over 4\kappa^2}={8\pi\rho_r\over a_r^4}t_r^2,\label{eq37}
\eeq
which is exactly the same as in GR. All of the above solutions also satisfy
the scalar field equation.

Much as with the dust case, these solutions
strongly resemble GR, due to cancellations of the scalar field terms.
If the scalars are constant during the radiation era, the gravitational
constant may be shifted. This may have observable consequences on
nucleosynthesis, for example. Deviations from exact Robertson-Walker
behavior, caused by primordial black holes or density fluctuations,
could also lead to regimes where the scalar fields play a more dynamic role.

\noindent{\bf cosmological constant era}

   When the universe is dominated by a cosmological constant, the matter is
given by
\beq
\rho = \rho_0 = {\rm constant},\hspace*{1cm} P=-\rho.\label{eq38}
\eeq
Such a situation arises, for example, in the inflationary
universe\cite{infl}, where $\rho_0$ is the energy density of a scalar field
either trapped in a false minimum or slowly rolling down a potential.
The $(00)$ equation again is similar to (\ref{eq26}), except that the
matter side is now given by $8\pi\rho_0$, and $F_d$ and $G_d$ are replaced
with the appropriate $F_0$ and $G_0$ in this regime. Examining the powers
of time, there are three, namely $-2$, $2q-2$, and 0. Unlike the previous
matter conditions, where the right hand side had $p$ dependence, these
three powers cannot in general be matched\footnote{There is one exception.
An exact solution can be found if the terms of power $2q-2$ in the $(00)$
equation cancel. In this case, equation (\ref{eq39}) is again valid, and
exponential solutions for the scale factor $a(t)$ and the $\phi$ are
found. By using the $(00)$ and $\phi$ equations, these conditions can all
consistently be met if $G_0=-4F_0$ and $q=H(-3\pm\sqrt{3})/2$. However, this
exact solution is really a special case of our first approximation,
(\ref{eq39}) - (\ref{eq42}),
when the terms ignored exactly cancel.}.
However, we may find
approximate solutions to the general case by assuming that one
of the two terms dominates gravity.

   If the fields are small, then the gravitational constant may dominate
the field contribution, $1/\kappa^2\gg F_{AB}\phi_A\phi_B$. Neglecting
terms involving the $\phi$ then gives the standard GR equation
\beq
3H^2 = 8\pi\kappa^2\rho_0,\label{eq39}
\eeq
which is solved by
\beq
a(t) = a_0\exp\left[\sqrt{\frac83\pi\kappa^2\rho_0}(t-t_0)\right].
\label{eq40}
\eeq
This exponential growth corresponds to an inflationary universe. The scalar
field equations give
\beq
6H^2F_{AB}\phi_B-G_{AB}(\ddot{\phi}_B+3H\dot{\phi}_B)=0.\label{eq41}
\eeq
Contracting with $\phi_A$ and solving the resulting equation gives
\beq
\phi_A=b_Ae^{q(t-t_0)}, \hspace*{1cm}
q={-3HG_0\pm\sqrt{9H^2G_0^2+24H^2F_0G_0}\over 2G_0}\label{eq42}
\eeq
as a solution. Thus, for the small field approximation, the $\phi$ may grow,
decay, or oscillate.

   If the $\phi$ are large compared to the gravitational constant, then a
power law ansatz for the growth of the scale factor and fields gives for
the $(00)$ equation
\beq
\left(3p^2F_0-\frac{q^2}2G_0+6pqF_0\right)t^{2q-2}=8\pi\rho_0 t_0^{2q}.
\label{eq43}
\eeq
Equating powers of time demands $q=1$, so the fields grow linearly,
and remain dominant over the GR term. Combining the field equations with
the time independent part of the $(00)$ equation gives two possibilities:
\barr
p = 0 &\Rightarrow& G_0 = -16\pi\rho_0 t_0^2\nn
p = \frac12 + {G_0\over 4F_0} &\Rightarrow&
{15\over4}F_0+\frac74G_0+{3\over16}{G_0^2\over F_0}=8\pi\rho_0 t_0^2.
\label{eq44}
\earr
The first case gives a static universe, while the second has the power of
expansion depending on the parameters of the theory.
For $p>1$, the second could give rise
to extended type inflation\cite{extend} if the phase transition is first
order, or soft inflation\cite{gets} if the potential is slowly rolling
of either the new or chaotic type.

   An exact solution can be found if the terms of power $2q-2$ in the $(00)$
equation cancel. In this case, equation (\ref{eq39}) is again valid, and
exponential solutions for the scale factor $a(t)$ and the $\phi$ are
found. By using the $(00)$ and $\phi$ equations, these conditions can all
consistently be met if $G_0=-4F_0$ and $q=H(-3\pm\sqrt{3})/2$. In fact, this
exact solution is really a special case of our first approximation,
when the terms ignored exactly cancel.

   In the inflationary scenario, the potential of a scalar field acts as
an effective cosmological constant to drive the expansion.
For $G_0>2F_0$, equation (\ref{eq44}) gives power law inflation. If the
potential is of the new\cite{new} or chaotic\cite{chaotic} type, then
this solution is a generalization of the soft inflationary
scenario\cite{gets}. In soft inflation, modifications of Einstein
gravity can remove the fine-tuning of potential parameters found in regular
gravity. The same situation can arise here, with the added advantage of
avoiding the solar system constraints.

   More interesting is if the potential is of the old inflation\cite{old}
type, giving rise to a first order phase transition. Simple nonminimal
coupling allows the phase transition to complete\cite{extend}. However,
to avoid an overproduction of big bubbles of true phase, which would
produce an excess of microwave background anisotropy, requires parameters
which violate the solar system constraint. Our theory can avoid this problem,
and can do it with parameters which exactly reproduce GR in the
solar system regime.
Unfortunately, as shown in the next section, the models which reproduce GR
are unstable in the dust era.

\setcounter{equation}{0}
\begin{flushleft}
\section{Two Field Case}
\end{flushleft}
\vspace{.5 cm}
\normalsize
\baselineskip = 24pt

   Studying the $N$ field case has the advantage not only of being general,
but also of allowing compact notation. However, the interrelationship of
parameters in these models can become obscured. We next consider the
simpler case of a two field model to better elucidate such features,
which include some stability problems.
The action for two fields, in analogy with equation (\ref{eq1}), is
\barr
S &=& \int d^4x\sqrt{-g}\left(f(\phi,\psi)R-{\omega\over2}(\nabla\phi)^2
-{\eta\over2}(\nabla\psi)^2+16\pi L_m\right),\label{eq45}\\
f(\phi,\psi) &=& {1\over\kappa^2}+h_1\phi^2+h_2\psi^2+h_3\phi\psi.\label{eq46}
\earr
A derivative cross term could also exist, but a proper rotation of the fields
will eliminate this term, so we set it to 0 without loss of generality.

   The solar system constraint, equation (\ref{eq15}), now becomes
\beq
(h_3^2\omega+4h_1^2\eta)\phi^2+4h_3(h_1\eta+h_2\omega)\phi\psi
+(h_3^2\eta+4h_2^2\omega)\psi^2=0,\label{eq46a}
\eeq
which may be rewritten as
\beq
\eta(h_3\psi+2h_1\phi)^2+\omega(h_3\phi+2h_2\psi)^2=0.\label{eq46b}
\eeq
Therefore, $\omega$ and $\eta$ must be opposite signs. We choose
$\omega$ to be positive. Furthermore, by rescaling the fields in the original
action, these kinetic constants may be chosen to be $\omega=1$, $\eta=-1$,
without any loss of generality. Thus, one important fact clearly shown in
the two field case is that multiple scalar tensor theories must have
negative kinetic terms if they are to agree with post-Newtonian tests
of general relativity.

{}From the relation (\ref{eq46b}), one can also see the interplay between
the kinetic
terms $\omega$ and $\eta$, the coupling to gravity through the $h_i$, and the
values of the fields themselves. Again, the asymptotic values of these
fields are determined by cosmology, and specifically, by their evolution
in the present dust era. We now study that evolution to derive further
constraints on the parameters of the theory.

   In a Robertson-Walker universe, the field equations become
\barr
(2h_1\phi+h_3\psi)(3\dot{H}+6H^2)-(\ddot{\phi}+3H\dot{\phi}) &=& 0\nn
(2h_2\psi+h_3\phi)(3\dot{H}+6H^2)+(\ddot{\psi}+3H\dot{\psi}) &=& 0.
\label{eq47}
\earr
Again, we make the assumption of power law behavior,
\beq
a(t)=a_c\left({t\over t_c}\right)^p,\hspace*{1cm}
\phi=b\left({t\over t_c}\right)^q,\hspace*{1cm}
\psi=c\left({t\over t_c}\right)^q.\hspace*{1cm}
\label{eq48}
\eeq
The field equations (\ref{eq47}) then combine to give
\beq
h_3={-2(h_1+h_2)bc\over b^2+c^2}.\label{eq49}
\eeq
Substituting this relationship back into the solar system constraint
(\ref{eq46b}) implies
\beq
b=\pm c,\label{eq50}
\eeq
which also yields the relationship between the $h_i$
\beq
h_3=\mp(h_1+h_2).\label{eq51}
\eeq

   These conditions also mandate that $f=1/\kappa^2$ in the dust era, as
noted earlier for $N$ fields. The $(00)$ and trace equations for the
two field case become
\barr
3H^2f+3H\dot{f}-\frac12(\dot{\phi}^2-\dot{\psi}^2)=8\pi\rho,\\
(6\dot{H}+12H^2)f+3\ddot{f}+9H\dot{f}+\dot{\phi}^2-\dot{\psi}^2
=8\pi(\rho-3P),
\earr
respectively. As in the $N$ field case, power law solutions exist.
In dust, the universe will expand with power $2/3$
just as in GR, and the same relation for the energy density,
(\ref{eq29}), still holds. The power $q$ at which the fields evolve is
\beq
q=\frac12\left(-1\pm\sqrt{1-\frac83(h_2-h_1)}\,\right).\label{eq52}
\eeq
The solutions for two fields in radiation and cosmological constant
dominated universes also proceed in analogous fashion, and are easily
obtainable from the more general $N$ field case by substituting the
appropriate expressions for $F$ and $G$.

\setcounter{equation}{0}
\begin{flushleft}
\section{Stability Analysis}
\end{flushleft}
\vspace{.5 cm}
\normalsize
\baselineskip = 24pt

   While the above solutions are exact, a crucial aspect of the system's
behavior is whether these solutions are stable. Only exceptionally
fine-tuned solutions will start with exactly the initial conditions to
fit the form of the above solutions. Further, any slight physical
fluctuation can move the system away from the exact solutions,
and the solar system constraints will be violated. Only
solutions which return to the exact behavior when deviating slightly
will likely satisfy the observable tests of these theories.
We therefore examine the evolution
of initially small perturbations about our power law solutions for the
two field case in the current dust universe.

   We start by writing the dust solutions for the case $b=c$, $h_3=-(h_1+h_2)$
as
\barr
a(t) &=& a_0\left({t\over t_0}\right)^{\frac23}[1+\epon(t)],\nn
\phi(t) &=& b\left({t\over t_0}\right)^q[1+\eptw(t)],\nn
\psi(t) &=& c\left({t\over t_0}\right)^q[1+\epth(t)],\label{eq53}
\earr
where $\epsilon_i\ll 1$ and $q$ is given by equation (\ref{eq52}) above.
The scalar field equations then become
\beq
\left[{4h_1\over3}-q(q+1)\right]
{\eptw\over t^2}-2(q+1){\dot{\eptw}\over t}-\ddot{\eptw}
-{2(h_1+h_2)\over3}{\epth\over t^2}
+\left[8(h_1-h_2)-3q\right]{\dot{\epon}\over t}
+3(h_1-h_2)\ddot{\epon}=0,\label{eq54}
\eeq
\beq
\left[-{4h_2\over3}-q(q+1)\right]
{\epth\over t^2}-2(q+1){\dot{\epth}\over t}-\ddot{\epth}
+{2(h_1+h_2)\over3}{\eptw\over t^2}
+\left[8(h_1-h_2)-3q\right]
{\dot{\epon}\over t}+3(h_1-h_2)\ddot{\epon}=0,\label{eq55}
\eeq
while the trace and $(00)$ equations give
\barr
&&\left[(12q^2+6q+\frac43)(h_1-h_2)+2q^2\right]{\eptw-\epth\over t^2}
+\left[6(2q+1)(h_1-h_2)+2q\right]{\dot{\eptw}-\dot{\epth}\over t}\nn
&&+3(h_1-h_2)(\ddot{\eptw}-\ddot{\epth})
+{1\over b^2\kappa^2}\left({t\over t_0}\right)^{-2q}
\left[{4\epon\over t^2}+{16\dot{\epon}\over t}+6\ddot{\epon}\right]=0,
\label{eq56}
\earr
\beq
(6q^3+7q^2+2q){\eptw-\epth\over t}
+(3q^2+2q)(\dot{\eptw}-\dot{\epth})
+{1\over b^2\kappa^2}\left({t\over t_0}\right)^{-2q}
({4\epon\over t}+4\dot{\epon})=0.\label{eq56a}
\eeq

   First consider the case of GR, with $\phi=\psi=0$. The trace
equation then gives
\beq
{2\over t^2}\epon+{8\over t}\dot{\epon}+3\ddot{\epon}=0.\label{eq57}
\eeq
This equation is solved by
\beq
\epon={A_1\over t}+{A_2\over t^{2/3}},\label{eq58}
\eeq
with $A_1$ and $A_2$ constant. The perturbation to the scale factor decays
in time, and hence GR is stable, as expected.

   Now, consider the perturbations of the scalar fields. Subtracting
one scalar field equation from the other gives
\beq
\ddot{\eptt}+{2\over t}(q+1)\dot{\eptt}=0,\label{eq61}
\eeq
where $\eptt\equiv\eptw-\epth$ and
equation (\ref{eq52}) was used to remove the $\eptt$ term.
This equation is solved by
\beq
\begin{array}{ll}
\eptt=Bt^{-2q-1}=Bt^{\pm\sqrt{1-\frac83(h_2-h_1)}}~~~~~~ & q\neq -\frac12,\nn
\eptt=B\ln t~~~~~~ & q = -\frac12,\label{eq62}
\end{array}
\eeq
with $B$ a constant. For the $q=-\frac12$ case the perturbation grows
logarithmically, and the scalar fields are unstable. When $q\neq -\frac12$
there are two roots, one of which grows in time with positive power.
In general, the scalar fields will pick up a combination of these two roots,
with coefficients depending on initial conditions in the dust era. Thus,
the perturbation $\eptt$ will generally have a component which grows in
time, and hence the scalar fields are unstable.

   That the solutions are unstable is not a surprise, for one of the fields
has a negative kinetic term. Such kinetic terms cause instability by
allowing for infinitely many negative energy states when the system is
quantized. They also can lead to the classical instability observed here.
In addition, our power law solutions include the case where the fields are
zero far from the solar system,
so that these values too will be unstable to perturbations
as the universe evolves. Hence, while multiple ST theories can satisfy
the solar system constraints, the requisite values will be unstable,
and hence are unlikely to actually occur.

   There are possible modifications to evade this instability. One of the
motivations for multiple ST theories is that they arise in string theories
and other models which attempt to incorporate gravity with the other
forces. The action used here may be just an approximation of the full theory,
whose other terms stabilize the evolution.
Or, on a classical level, a mass or potential term for the scalars could
freeze the $\phi$ at some minimum, thus preventing the runaway growth of
small perturbations. Any such modification would have to be examined in
more detail to see if stability arises. However, given the problems
associated with negative kinetic terms, multiple ST theories
that are consistent with GR in the weak-field limit seem to be unstable.

\setcounter{equation}{0}
\begin{flushleft}
\section{Conclusion}
\end{flushleft}
\vspace{.5 cm}
\normalsize
\baselineskip = 24pt

   Multiple scalar ST theories offer the possibility of providing a new
theoretical framework in which to test GR. Because such theories can
mimic GR in the solar system, where regular BD theories are severely
limited, but yield substantially different results in other regimes,
they are especially valuable.
Our results complement previous work\cite{Nfield}, which
used values constant in time for the scalar fields far from massive
compact objects. The authors of this previous work
also note the need for negative kinetic terms, and point out the
potential for problems with quantum and stellar stability. We have
found still another problem.
In theories like this, with negative kinetic energy terms,
the cosmological solutions which meet the solar
system constraints are
unstable as they evolve in time, and thus can not be maintained.

Our time evolving solutions for the scalars are
consistent with both solar system tests of GR and the cosmological
field equations.
If these solutions can be stabilized by some means such
as an explicit potential term or a more complete theory of gravity,
then they will be a useful probe of GR.
However, without such stabilization, the only way to make these theories
agree with GR in the solar system is to choose constants that make the
theory approximate GR not only in the post-Newtonian approximation, but
also in other regimes. This situation is in analogy with the single
scalar JBD theory, and going to multiple scalars would have no advantage.

   Finally, we note that our solution in the cosmological
constant dominated regime exhibits power law growth when the fields
dominate, as seen in equation (\ref{eq44}). For power greater than 1,
this yields inflation. The extended inflation scenario\cite{extend} used
the JBD theory
to create power law inflation, in which case the first order phase
driving inflation would eventually complete, in contrast with the GR
case. However, the JBD theory was inadequate, because values of the JBD
constant which met the solar system constraint produced too many big
bubbles, and thus an inhomogeneous universe\cite{wbg}.
One might hope to use multiple scalar ST theories to give both power law
inflation as well as meeting both the solar system constraint and the big
bubble constraint. However, this scenario will not likely occur, as the
dust values of the fields are unstable. Should a stabilizing mechanism be
found, then further investigation of the inflationary scenario in these
models would be warranted.

\begin{flushleft}
{\bf Acknowledgements}
\end{flushleft}

We thank Kirk Brattkus, Kei-ichi Maeda, Larry Romans, Bahman Shahid-Saless,
and Eric Woolgar for useful conversations.
ALB acknowledges support from the National Research Council.
The research described in this publication was
carried out by the Jet Propulsion Laboratory, California Institute of
Technology, under a contract with the National Aeronautics and Space
Administration.

\end{document}